\documentclass[%
 aip,
 amsmath,amssymb,
 reprint,%
]{revtex4-1}

\usepackage{graphicx}% Include figure files
\usepackage{dcolumn}% Align table columns on decimal point
\usepackage{bm}% bold math
\usepackage[utf8]{inputenc}
\usepackage[T1]{fontenc}
\usepackage{mathptmx}
\usepackage{placeins}
\usepackage{upgreek}

\newcommand{\umdot}{$\upmu\mathrm{m}$}

\usepackage{xcolor} %So we can write red highlighted things

\usepackage{hyperref} %So we can standardize the subfigure calls
\newcommand*{\figref}[2][]{%
  Fig.~\hyperref[{#2}]{%
    \ref*{#2}%
    \ifx\\(#1)\\%
    \else
      (#1)%
    \fi
  }%
}

\begin{document}

\preprint{AIP/123-QED}

\title[Quantum networks based on color centers in diamond]{Quantum networks based on color centers in diamond}

\vspace{3cm}

\author{Maximilian Ruf}
\affiliation{QuTech, Delft University of Technology, PO Box 5046, 2600 GA Delft, The Netherlands}
\affiliation{Kavli Institute of Nanoscience Delft, Delft University of Technology, PO Box 5046, 2600 GA Delft, The Netherlands}

\author{Noel H. Wan}
\affiliation{Department of Electrical Engineering and Computer Science, Massachusetts Institute of Technology, Cambridge, Massachusetts 02139, USA}
\affiliation{Research Laboratory of Electronics, Massachusetts Institute of Technology, Cambridge, Massachusetts 02139, USA}

\author{Hyeongrak Choi}
\affiliation{Department of Electrical Engineering and Computer Science, Massachusetts Institute of Technology, Cambridge, Massachusetts 02139, USA}
\affiliation{Research Laboratory of Electronics, Massachusetts Institute of Technology, Cambridge, Massachusetts 02139, USA}

\author{Dirk Englund}
\email{englund@mit.edu} 
\affiliation{Department of Electrical Engineering and Computer Science, Massachusetts Institute of Technology, Cambridge, Massachusetts 02139, USA}
\affiliation{Research Laboratory of Electronics, Massachusetts Institute of Technology, Cambridge, Massachusetts 02139, USA}
\affiliation{Brookhaven National Laboratory, Upton, New York 11973, USA}

\author{Ronald Hanson}
\email{r.hanson@tudelft.nl} 
\affiliation{QuTech, Delft University of Technology, PO Box 5046, 2600 GA Delft, The Netherlands}
\affiliation{Kavli Institute of Nanoscience Delft, Delft University of Technology, PO Box 5046, 2600 GA Delft, The Netherlands}

\date{\today}

\begin{abstract}
With the ability to transfer and process quantum information, large-scale quantum networks will enable a suite of fundamentally new applications, from quantum communications to distributed sensing, metrology, and computing. This perspective reviews requirements for quantum network nodes and color centers in diamond as suitable node candidates. We give a brief overview of state-of-the-art quantum network experiments employing color centers in diamond, and discuss future research directions, focusing in particular on the control and coherence of qubits that distribute and store entangled states, and on efficient spin-photon interfaces. We discuss a route towards large-scale integrated devices combining color centers in diamond with other photonic materials and give an outlook towards realistic future quantum network protocol implementations and applications.
\end{abstract}

\maketitle

\begin{figure*}
\includegraphics[width=0.99\textwidth]{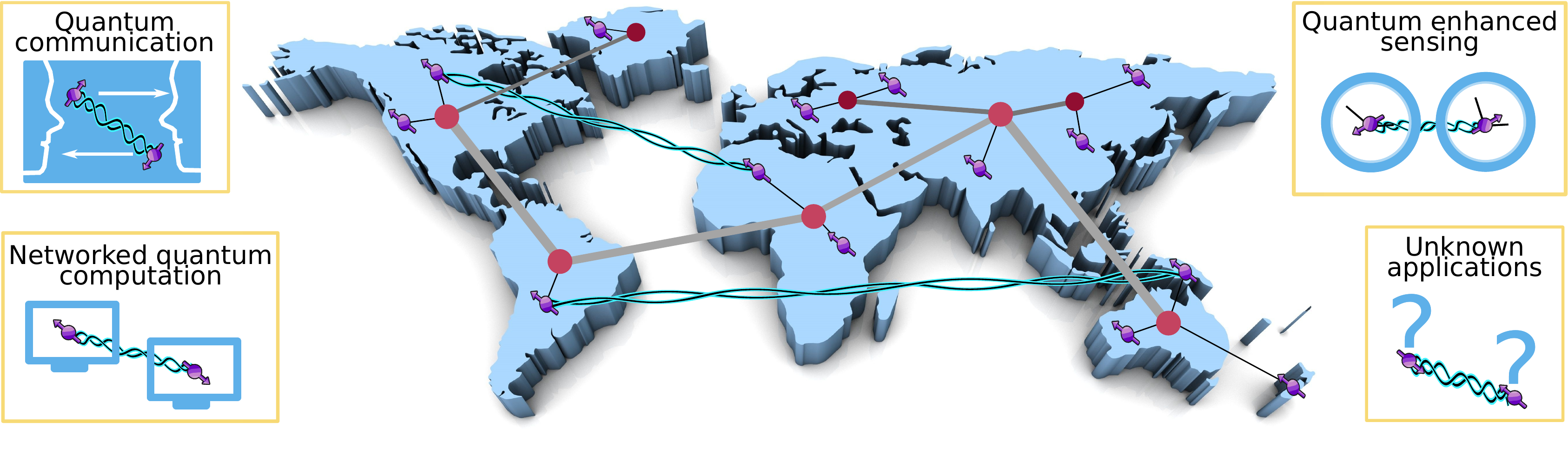}
\caption{Schematic overview of a future large-scale quantum network, consisting of nodes containing optically interfaced qubits (purple) with long coherence times. Photons routed via optical fibers or free-space channels serve as mediators to create entanglement (blue wiggled lines). Local area and trunk backbone quantum repeaters (dark and light red circles, respectively) are used to enable a high entanglement generation rate over large distances, overcoming photon transmission loss. The entanglement generation is heralded, meaning that detection of certain photon statistics signals the successful generation of an entangled state that is available to a network user for further processing and applications such as networked quantum computation, quantum secured communication, quantum enhanced sensing, and potential not yet discovered tasks.}
\label{fig:quantum_network_overview}
\end{figure*}

 With the Panama-Pacific International Exposition of 1914 fast approaching, AT\&T's leadership grew increasingly alarmed: it was still impossible to hold a coast-to-coast telephone call, despite  the company touting it for years. The essential problem ``was a satisfactory telephone repeater,'' recalled an AT\&T senior manager from conversations with the company's chief engineer John J. Carty~\cite{Gertner2012}. Whereas mechanical repeaters adequately boosted voice signals on the metro-scale, they added so much noise that signals quickly became unintelligible over longer distances. A fundamentally new repeater technology was needed. Tasked with this urgent mission, Bell Laboratories developed the ``audion'' trion vacuum tube, just in time for the 1914 Exposition. Over the ensuing decades, the triode made way to the transistor, copper wires yielded to optical fiber, and binary digits (''bits") became the universal language of information, transcending physical modality. Today, as the authors edit these sentences over a globe-spanning video conference, the world is at the cusp of the next information revolution, as quantum bits (''qubits") have become the universal units fueling a new generation of `quantum information technologies'. And once again, an essential challenge is to develop a ``satisfactory repeater'' -- this time, capable of relaying (but not amplifying!~\cite{Park1970}) \textit{quantum information} signals. This article reviews the progress towards one such ``quantum repeater'' as well as other quantum network technology using color centers in diamond to connect quantum information among spins and photons.

\section{A brief introduction to quantum networks}

In a future quantum network (see Fig.~\ref{fig:quantum_network_overview}), remote parties are connected by sharing long-lived entangled states~\cite{Kimble2008,Wehner2018}. Arguably, the most promising way of linking distant nodes is to employ fiber- or free-space photonic communication channels to establish entanglement. While all photon-based schemes are associated with losses that scale with distance~\cite{Pirandola2017}, motivating the need for quantum repeaters~\cite{Briegel1998}, heralding entanglement generation on successful photon transmission events maps these losses into reduced entanglement generation rates without lowering entanglement fidelities~\cite{Cabrillo1999,Barrett2005,Duan2004}. 

Optically-mediated remote entanglement of individually controllable qubits has been generated for different materials platforms, including quantum dots~\cite{Delteil2016,Stockill2017}, trapped ions~\cite{Moehring2007,Hucul2015,Stephenson2020}, neutral atoms~\cite{Ritter2012,Hofmann2012,Daiss2021,Langenfeld2021}, and nitrogen-vacancy centers in diamond~\cite{Bernien2013,Hensen2015}. Other promising systems, including so-called group-IV defects in diamond ~\cite{Bhaskar2017,Nguyen2019,Bhaskar2020,Trusheim2020}, defects in SiC~\cite{Nagy2019,Bourassa2020,Lukin2020}, and rare-earth ions in solid-state hosts~\cite{Raha2020,Kindem2020,Merkel2020}, show great potential for quantum network applications, although remote entanglement has not yet been generated. Another less explored approach is to link distant superconducting quantum processors using coherent conversion of microwave photons to telecom frequencies~\cite{Chu2020,Mirhosseini2020,Forsch2020,Hease2020, Krastanov2020}.

Apart from fundamental tests of physics~\cite{Hensen2015}, small-scale quantum networks have been used to demonstrate key network protocols such as non-local quantum gates~\cite{Daiss2021}, entanglement distillation~\cite{Kalb2017}, and very recently entanglement swapping~\cite{Pompili2021}. These networks are currently limited to a few nodes~\cite{Pompili2021}, distances of up to one kilometer~\cite{Hensen2015}, and entanglement generation rates in the Hz to kHz regime~\cite{Delteil2016,Stockill2017,Moehring2007,Hucul2015,Stephenson2020,Ritter2012,Hofmann2012,Bernien2013,Hensen2015}. A major challenge for the coming decade is to transition from the current proof-of-principle experiments to large-scale quantum networks for use in fields such as distributed quantum computation~\cite{Nickerson2014,choi2019percolation}, quantum enhanced sensing~\cite{Gottesman2012,Komar2014}, and quantum secure communication~\cite{Ekert2014}. 

In this perspective article, we focus on color centers in diamond as potential building blocks for large-scale quantum networks. We identify their strengths and challenges, current research trends, and open questions, with a focus on qubit control, coherence, and efficient spin-photon interfaces. We lay out a path towards large-scale, integrated devices, and discuss possible protocols and functionalities such devices could enable. 

This article is organized as follows:  Sec.~\ref{Sec.:requirements_quantum_node} introduces the requirements for the nodes of a quantum network and gives a high-level overview of why color centers in diamond are promising candidates to meet these demands. We then give a brief overview of state-of-the-art quantum network experiments employing color centers in diamond in Sec.~\ref{Sec.:state_of_the_art_qns}, to give the reader a sense of the current research status. In Sec.~\ref{Sec.:Optial_and_spin_properties}, we introduce the physics behind color centers in diamond, to understand their strengths and challenges for quantum network applications in the remainder of this article. Sec.~\ref{Sec.:enhancing_memories} discusses open questions regarding the coherence and control of nuclear spin memory qubits surrounding color centers in diamond that can be used as additional quantum resources. In Sec.~\ref{Sec.:enhancing_optical_interface}, we outline the need and strategies for an enhancement of the spin-photon interface of diamond color centers and discuss promising future research in this direction. Sec.~\ref{Sec.:Large-scale integrated devices} outlines a path towards large-scale, integrated diamond devices that may enable future quantum network experiments with high rates. Finally, Sec.~\ref{Sec.:Conclusions} concludes by illuminating possible applications for future quantum networks.

\section{Requirements for a quantum network node} \label{Sec.:requirements_quantum_node}

\begin{figure*}[t!]
\includegraphics[width=.7\textwidth]{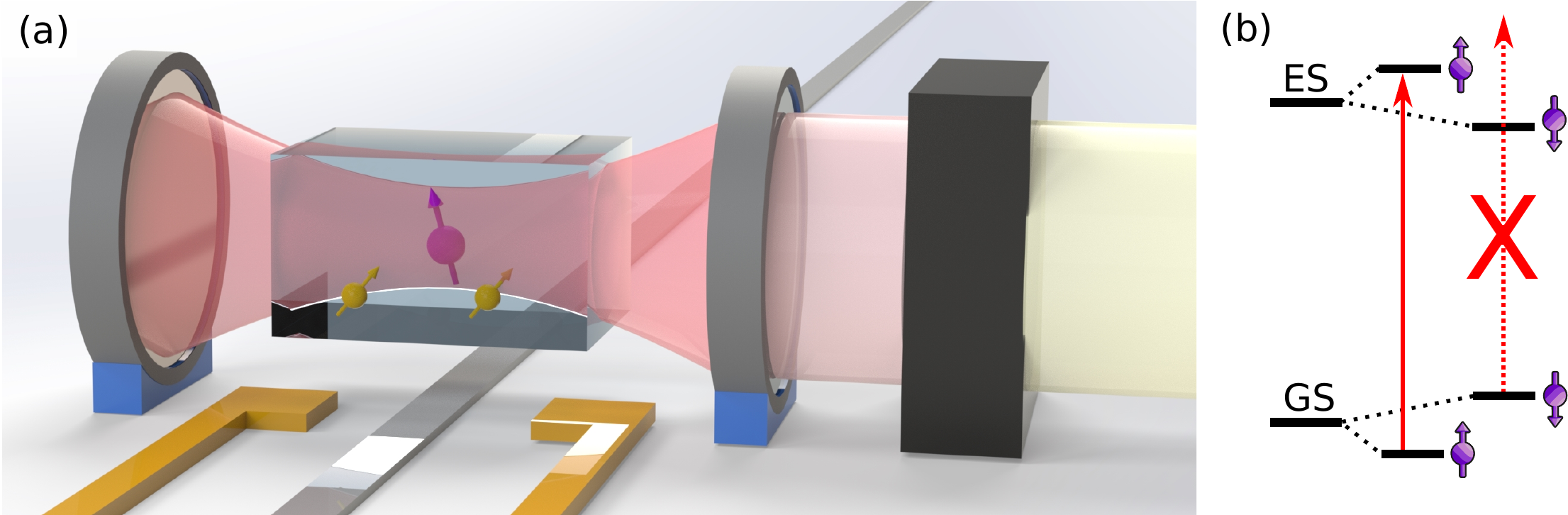}
\caption{Overview of a quantum network node based on color centers in diamond and simplified optical level scheme. (a) Schematic of a quantum network node in diamond, consisting of an optically-active (communication) qubit (purple) embedded between two highly reflective mirrors (an optical cavity) to enhance the interaction strength of photons with the qubit. The state of the color center qubit can be swapped onto long-lived nuclear spin (memory) qubits in the surroundings (orange) using high fidelity gates that employ microwave or RF pulses delivered via dedicated lines (gray). Tuning of network nodes to an optimal frequency operation point is e.g.~enabled by applying a static electric field via electrodes (yellow).  Frequency-conversion (black box) can be used to down-convert photons entangled with the state of the communication qubit to telecommunication wavelengths, for which photon transmission losses are low. The color center features spin-resolved optical transitions between an optical ground (GS) and excited state (ES), enabling optically-mediated remote entanglement generation (b).}
\label{fig:quantum_network_node_schematic}
\end{figure*}

To make a material platform suitable for a node in a quantum network in which entanglement is mediated by photons, it has to fulfill three main requirements~\cite{Wehner2018}. First, the capability to interface at least one qubit efficiently with optical photons (at telecommunication wavelengths for fiber-based systems), to establish remote entanglement at high rates. Second, the ability to store quantum states during entanglement generation; in particular, this requires qubit coherence times under full network activity to be longer than the time it takes to generate entanglement between nodes. Third, the capability to store several entangled states per node with a capability for high-fidelity operations between them, to enable multi-qubit protocols such as error correction.

Color centers in diamond satisfy most of these requirements, and have enabled some of the most advanced experimental demonstrations of quantum network protocols to date. We here give a concise, high level overview of the achieved and projected key capabilities of color centers in diamond [see \figref[a]{fig:quantum_network_node_schematic}], motivating a detailed treatment (and also including the relevant literature) in the remains of this article:

\begin{enumerate}
    \item The color center contains an individually controllable, optically active spin (communication qubit), with access to several long-lived nuclear spins (memory qubits) in its surrounding. These memory qubits can be manipulated with high fidelity to free up the communication qubit and enable multi-qubit protocols. They have a long coherence time enabling robust state storage during subsequent network activity. 

    \item The internal level structure is suited to generate remote entanglement. In particular, color centers feature spin-state-selective optical transitions that can entangle the color center's spin state with a photonic state, e.g.~in the number / polarization / time basis (see \figref[b]{fig:quantum_network_node_schematic}.)

    \item Optical emission of color centers is bright and can be collected with high efficiency, enabling high entanglement generation rates. This emission is in the visible to near-IR spectrum, and thus at wavelengths that are associated with higher fiber transmission losses than for photons in the telecom band. However, it is possible to efficiently convert this emission to the telecom band while maintaining quantum correlations.

   \item Diamond is a solid-state material, potentially enabling nanophotonic device fabrication at a large scale. Such devices can also be integrated with photonic circuits, and efficient interfacing of color centers embedded in such systems has been demonstrated.

\end{enumerate}

\section{State of the art of diamond-based quantum networks} \label{Sec.:state_of_the_art_qns}

\begin{figure}[t!]
\includegraphics[width=.49\textwidth]{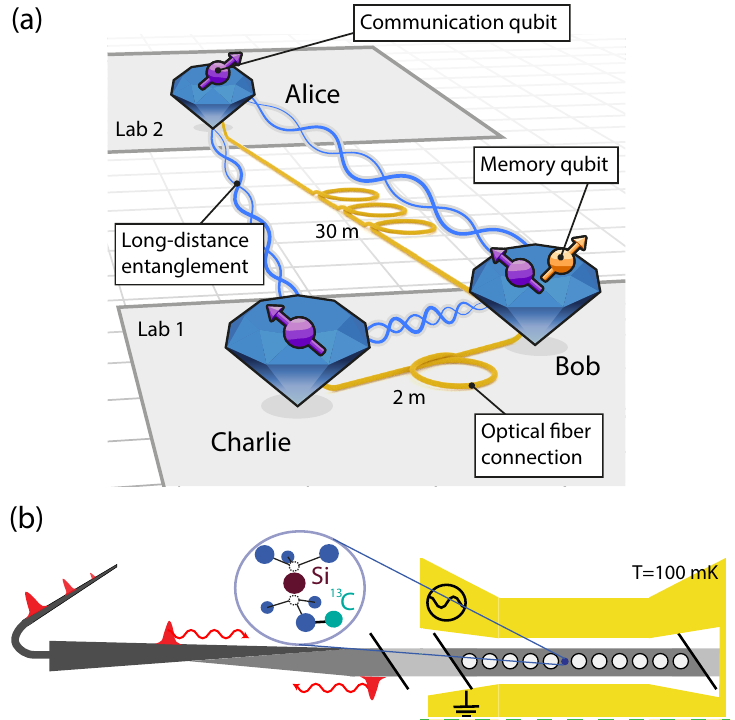}
\caption{State-of-the-art experiments showing the potential of color centers in diamond as quantum network nodes. (a) Schematic of a 3-node quantum network based on nitrogen-vacancy (NV) centers in diamond, linked with optical fibers. One of the network nodes features a $^{13}$C nuclear spin that is used as a memory qubit to store (part of) an entangled state and free up the communication qubit, while subsequent entanglement is created using this communication qubit a second time. Figure from ~\cite{Pompili2021}, reprinted with permission from AAAS. (b) Schematic of a silicon-vacancy (SiV) center (blue circle) embedded in a fiber-coupled (dark grey) all-diamond photonic crystal cavity (light grey area with holes). Coherent emitter-cavity cooperativities $\sim 100$ have enabled high-fidelity reflection-based spin-photon entanglement generation~\cite{Bhaskar2020}, and control of individual $^{13}$C nuclear spins via microwave pulses applied to the electron spin of the silicon vacancy spin qubit (yellow electrodes) has been demonstrated~\cite{Nguyen2019overview}. Figure adapted with permission from Ref.~\cite{Nguyen2019overview}. Copyright (2019) by the
American Physical Society.}
\label{fig:quantum_networks_state_of_the_art}
\end{figure}

Two recent experiments showcasing the potential of color centers in diamond as quantum network node candidates can be seen in Fig.~\ref{fig:quantum_networks_state_of_the_art}, together defining the state of the art in quantum networks. \figref[a]{fig:quantum_networks_state_of_the_art} shows a laboratory-scale quantum network based on nitrogen-vacancy (NV) centers in diamond \cite{Pompili2021}. Three NV centers in separate cryostats, linked via optical fiber channels in a line configuration, are operated as independent quantum network nodes. The centers are set to a common optical emission frequency via electric field tuning (DC Stark shift)~\cite{Bernien2013}. The network is used to demonstrate distribution of three-partite Greenberger-Horne-Zeilinger (GHZ) entangled states across the nodes, as well as entanglement swapping to achieve any-to-any connectivity in the network~\cite{Pompili2021}. This network sets the state of the art for entanglement-based quantum networks.

Entanglement between the nodes is generated using a photon-emission-based scheme, in which spin-photon entanglement is locally generated at each NV node using microwave and laser pulses. The photonic modes are then interfered on a beam splitter, thereby erasing the which-path information. Subsequent photon detection heralds the generation of a spin-spin entangled state between two NV nodes~\cite{Bernien2013,Humphreys2018}. Early experiments~\cite{Bernien2013} had employed two-photon entangling schemes reaching fidelities $\sim 90 \%$~\cite{Hensen2015} at $\sim$ mHz rates. More recent experiments (including the three-node network discussed here) used a single-photon scheme~\cite{Humphreys2018} yielding fidelities $\sim 82 \%$ at rates of $\sim 10$ Hz\cite{Pompili2021}.

Importantly, these schemes generate entanglement in a heralded fashion, making the fidelity robust to photon loss and the generated entangled states available for further use. In the vicinity of the NV center, a $^{13}$C nuclear spin is used as a memory qubit at the middle node, coupled to the NV center electron spin via the hyperfine interaction. The nuclear spin is controlled via tailored microwave pulses applied to the NV center spin~\cite{Taminiau2014} that simultaneously decouple the NV electron spin from the other nuclear spins. Characterization of the nuclear spin environment of an NV center has enabled communication qubit coherence times above one second~\cite{Abobeih2018}. In addition, entangled states have been stored in $^{13}$C memory qubits for up to $\sim 500$ subsequent entanglement generation attempts~\cite{Reiserer2016,Kalb2018,Pompili2021}. Combined with the demonstrated capability of entanglement distillation (generating one higher fidelity state from two lower fidelity states)~\cite{Kalb2017} and deterministic entanglement delivery (by generating entanglement between two nodes faster than it is lost, albeit not under full network activity)~\cite{Humphreys2018}, these experiments highlight the potential of the NV center as a quantum network node. As we discuss in detail below, a main challenge for the NV center is the low fraction of emitted and detected coherent photons, limiting the entanglement rates.

Recently, a new class of defect centers in diamond has also gathered the attention of the community: the family of group-IV color centers in diamond, of which the negatively charged silicon-vacancy (SiV) center is the most studied member. \figref[b]{fig:quantum_networks_state_of_the_art} shows a schematic of a SiV center embedded in a photonic crystal cavity, which increases the interaction of light with the communication qubit as compared to emission in a bulk diamond material (discussed in Sec.~\ref{Sec.:enhancing_optical_interface} below). Spin-photon entanglement with fidelity $\geq$ 94 \% has been generated in this system using cavity quantum electrodynamics (cQED) reflection-based schemes, in which photons impinging on the cavity containing the communication qubit are reflected or transmitted depending on the spin-state of the communication qubit~\cite{Nguyen2019overview}. While entanglement between distant SiV centers has not yet been demonstrated, strong collective interactions between two SiV centers~\cite{Evans2018} in the same cavity and probabilistic entanglement generation between two SiVs ~\cite{Sipahigil2016,Machielse2019} in one waveguide have been observed. In addition, photon outcoupling efficiencies from the emitter-cavity system into fiber $>$ 0.9 have enabled the demonstration of memory-enhanced quantum communication~\cite{Bhaskar2020}. This experiment sets the state of the art for quantum repeater experiments.

Additionally, recent experiments have demonstrated key quantum network node capabilities, including dynamical decoupling of the SiV spin from the bath with coherence times above 1 ms~\cite{Sukachev2017,Nguyen2019}, $^{13}$C nuclear spin initialization, control, and readout via the central SiV electron spin~\cite{Metsch2019,Nguyen2019detail}, and frequency control of SiV centers via strain tuning~\cite{Sohn2018,Machielse2019,Wan2020}. Combined with recent efforts to fabricate SiV centers in nanophotonic waveguides at a large scale and their integration in photonic circuits~\cite{Wan2020}, these results show the potential of silicon-vacancy centers and other closely related group-IV color centers as nodes for a quantum network (see Sec.~\ref{Sec.:enhancing_optical_interface}).

\section{Optical and spin properties of color centers in diamond} \label{Sec.:Optial_and_spin_properties}

\begin{figure}
\centering
\includegraphics[width=.475\textwidth]{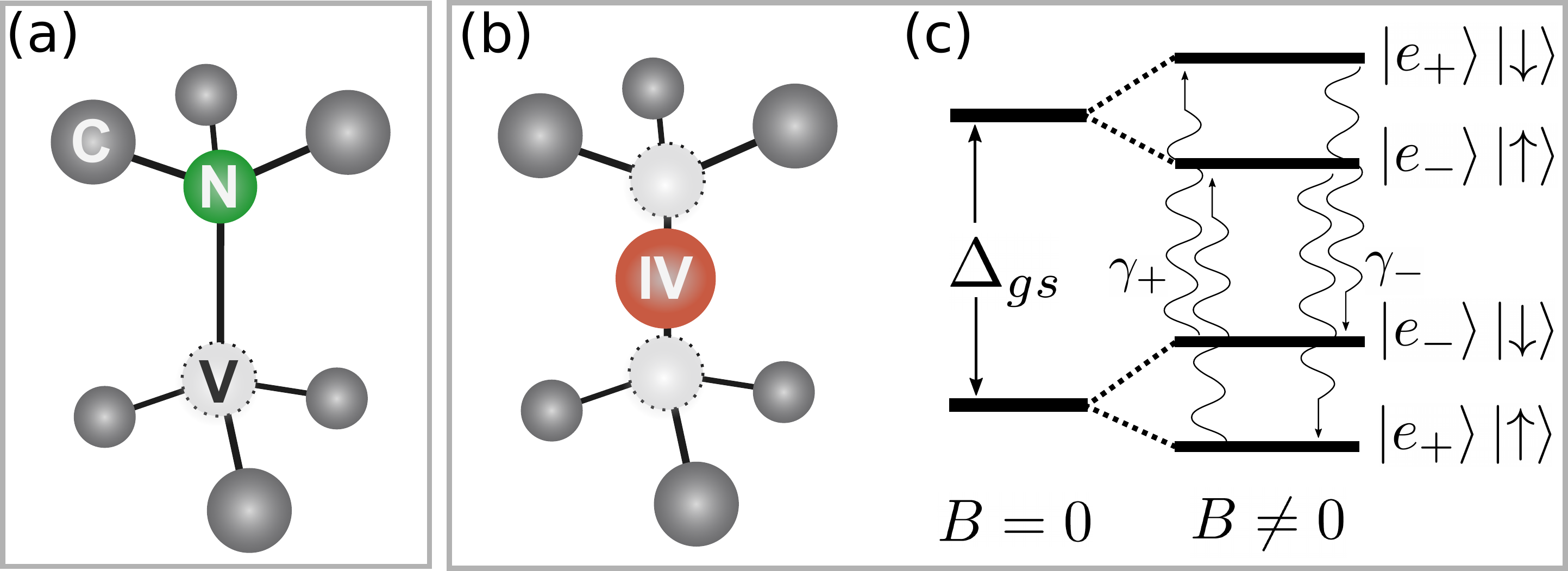}
\caption{Schematic atomic structure of NV (a) and group-IV (b) color centers in diamond and ground state level structure of group-IV color centers (c). The inversion symmetry of group-IV color centers leads to a vanishing permanent electric dipole moment, making those defect centers suitable for integration in nanophotonic structures. This structure also results in orbital and spin double degeneracy in the ground and excited state (c, only ground state shown for simplicity) that is lifted by (mostly) spin-orbit coupling ($\Delta_{gs}$) and by application of an external magnetic field ($B$), respectively. Phonon-induced transitions ($\gamma_{+}$ / $\gamma_{-}$) between orbital states limit (communication) qubit coherence times (see text). }
\label{fig:snv_nv_schematic}
\end{figure}

Nitrogen-vacancy and group-IV color centers in diamond feature different structural symmetries, which result in different properties of their spin-photon interface. Here, we briefly introduce the most important differences between those color centers in diamond, to the point necessary to understand current limitations and research directions described in the remains of this article; we refer the reader to the many excellent articles that cover the physics of these color centers in great detail~\cite{Doherty2013,Hepp2014thesis,Bradac2019}. In this article, we consider all color centers to be in the negative charge state, except where indicated differently; this negative charge state is the one most commonly studied and used in experiments geared towards entanglement generation.

\begin{table*}
\centering
\begin{tabular}{ccccccc}
Vacancy     &   ZPL      & Quantum        & Debye-Waller      &  Radiative   & Ground state splitting & $T_{\textrm{oper}}$ \\
     &   wavelength [nm]     & efficiency [\%]       & factor       &  Lifetime [ns] & [GHz] & [K]\\ [0.12cm]\hline \hline
SiV     &    737~\cite{Bhaskar2020}    & $(1-10)$~\cite{Neu2012,Sipahigil2016} / $ 60$~\cite{Riedrich-Moller2014} & $ (0.65-0.9)$~\cite{Neu2011,Dietrich2014}  &  $ 1.7$~\cite{Rogers2014,Sipahigil2016} &   $ 50$~\cite{Hepp2014,Rogers2014,Pingault2017}  & 0.1 \\ [0.12cm]
GeV     &   $ 602$~\cite{Iwasaki2015}    & $ (12-25)$~\cite{Jensen2019,Nguyen2019b}       & $ \sim0.6$~\cite{Siyushev2017}  & $ (1.4-6)$~\cite{Iwasaki2015,Bhaskar2017} & $ 170$~\cite{Siyushev2017} &  0.4\\ [0.12cm]
SnV     &   $ 620$~\cite{Iwasaki2017, Trusheim2020}    & $ \sim80$~\cite{Iwasaki2017}       & $ 0.57$~\cite{Gorlitz2019}  & $ (4.5-7)$~\cite{Tchernij2017,Trusheim2020,Rugar2021} &   $ 850$~\cite{Iwasaki2017,Trusheim2020} & 1.8\\ [0.12cm]
PbV     &   $ 520$~\cite{Trusheim2019}    & not known         & not known  & $> 3$~\cite{Trusheim2019} &   5700~\cite{Trusheim2019} & 9.8\\ [0.12cm]\hline
NV      &   $ 637$~\cite{Doherty2013}    & $> 85$~\cite{Radko2016}     & $ 0.03$~\cite{Faraon2011,Riedel2017a} &  $ (11-13)$~\cite{Faraon2011,Kalb2018,Ruf2021} & not applicable & not applicable \end{tabular}
\caption{Overview of optical and orbital properties of group-IV (SiV, GeV, SnV, PbV) and NV defect centers in diamond. All listed numbers are approximate values. We consider all defects to be in the negative charge state. The Debye-Waller Factor characterizes the fraction of radiative emission into the ZPL, while we define the quantum efficiency as the ratio of radiative to all decay. Ground state splitting denotes splitting between orbital states in the optical ground state of group-IV color centers (see \figref[c]{fig:snv_nv_schematic} and text). $T_{\textrm{oper}}$ is the calculated operation temperature below which the expected orbital lifetime is above 100 ms. We estimate this temperature by calculating the phonon transition rate $\gamma_{+} \propto \Delta_{gs}^3/(e^{(\Delta_{gs/k_bT)}}-1)$ [see \figref[c]{fig:snv_nv_schematic}], where $\Delta_{gs}$ is the orbital ground state splitting, $T$ is the temperature, and $k_B$ the Boltzmann constant (as $\gamma_{-}$ > $\gamma_{+}$, $\gamma_{+}$ determines the qubit coherence times), as in Refs.~\cite{Iwasaki2017,Trusheim2020}. We then find the proportionality factor to map this to a concrete temperature by using the measured values of Ref.~\cite{Pingault2017}.}
\label{tab:nvsnvproperties}
\end{table*}

Fig.~\ref{fig:snv_nv_schematic} shows the difference in atomic structure of nitrogen-vacancy (NV) centers and group-IV color centers (consisting of silicon-vacancy (SiV)~\cite{Hepp2014,Pingault2017,Nguyen2019detail}, germanium-vacancy (GeV)~\cite{Siyushev2017}, tin-vacancy (SnV)~\cite{Iwasaki2017,Trusheim2020} and lead-vacancy (PbV)~\cite{Trusheim2019} centers). An NV center is formed by a nitrogen atom replacing a carbon atom of the diamond lattice, next to a lattice vacancy, giving rise to a C$_{3\textrm{v}}$ symmetry [\figref[a]{fig:snv_nv_schematic}]. The NV center possesses a permanent electric dipole moment, making the optical transition frequencies sensitive to charge fluctuations in its environment~\cite{Doherty2013}. This sensitivity to electric fields has been used as a resource to tune different NV centers onto resonance~\cite{Bernien2012}, counteracting local strain inhomogeneities in the host diamond crystal which shift the transition frequencies of color centers. However, it has also hindered the nanophotonic integration of NV centers due to degradation of their optical properties near surfaces~\cite{Faraon2012,Ruf2019}.

The NV center features an outstanding spin energy relaxation time of over 8 h~\cite{Astner2018}. Quantum states have been stored in the NV center's electronic spin for over one second by decoupling the electron from unwanted interactions in its environment using tailored microwave pulse sequences~\cite{Abobeih2018}. Fidelities in quantum networks experiments have reached $>0.99\%$ for single qubit gates and $\sim 98 \%$ for two-qubit gates (between NV electron and $^{13}$C nuclear spins) ~\cite{Pompili2021}.

For group-IV color centers, the group-IV atom takes an interstitial lattice site between two lattice vacancies. The resulting inversion-symmetric D$_{3_d}$ structure of the defect [\figref[b]{fig:snv_nv_schematic}] results in no permanent electric dipole and thus a first-order insensitivity to electric field fluctuations. As a consequence, tuning optical transition frequencies of group-IV color centers by electric fields is relatively inefficient~\cite{DeSantis2021,Aghaeimeibodi2021}, potentially necessitating other techniques (discussed in Sec.~\ref{Sec.:enhancing_optical_interface} below). On the other hand, the vanishing permanent electric dipole makes these color centers excellent candidates for integration into photonic nanostructures (also discussed in Sec.~\ref{Sec.:enhancing_optical_interface} below). 

Group-IV emitters also feature a different level structure than the NV center: both optical ground and excited states are formed by doubly-degenerate orbital states, which are each split into pairs by spin-orbit interaction and the Jahn-Teller effect. Under an external magnetic field, the double spin degeneracy in each pair ($S = 1/2$) is lifted~\cite{Hepp2014,Hepp2014thesis,Thiering2018}. The qubit subspace can then for example be defined over the two opposing spin sublevels of the optical ground state's lower orbital branch [see \figref[c]{fig:snv_nv_schematic}]. This, however, means that direct microwave driving of the qubit levels is forbidden in first order, as one would need to flip both spin and orbital quantum numbers. Typically, this complication is overcome by working with emitters with transversal strain $\sim \Delta_{gs}$, which mixes the orbital levels, such that the qubit state can be effectively described by the spin only and thus be driven with higher efficiency~\cite{Nguyen2019detail}. Coherent microwave control of group-IV color centers has recently been demonstrated for SiV centers~\cite{Sukachev2017,Nguyen2019detail}, with infidelities on the percent level~\cite{Sukachev2017}. For other group-IV color centers, only incoherent microwave driving of spin transitions has been achieved so far~\cite{Siyushev2017,Trusheim2020}. The reported infidelities for coherent SiV control are a result of qubit dephasing during microwave pulses, as relatively low Rabi frequencies ($\sim$ 10 MHz) are used to limit heating-induced decoherence. It is expected that the use of superconducting waveguides~\cite{Sigillito2014} will further improve microwave pulse fidelities (as a result of potentially significantly reduced ohmic heating) for all color centers. 

Spin dephasing in the group-IV qubit subspace is dominated by phonon-assisted transitions to the upper ground state orbital branch, which leads to the acquisition of a random phase: while the phonon transitions are in principle spin-conserving, the detuning between the spin states varies depending on which orbital state is occupied, see \figref[c]{fig:snv_nv_schematic}~\cite{Jahnke2015,Meesala2018}. In Tab.~\ref{tab:nvsnvproperties}, we map reported ground state splittings into a concrete performance metric by extrapolating the temperatures $T_{\textrm{oper}}$ below which we expect an orbital coherence time $> 100$ ms due to low phonon occupation. From this, we see that a larger ground state splitting (associated with stronger spin-orbit coupling strengths for heavier ions) is beneficial for maintaining long qubit coherence. However, this requires higher microwave driving powers for similar off-axis strain values, as the strain-induced orbital mixing is related to the spin-orbit coupling strength, and thus the overlap between orbital states is reduced~\cite{Nguyen2019detail}. All-optical spin control schemes represent an alternative to microwave driving, as demonstrated e.g.~for SiV~\cite{Rogers2014,Pingault2014,Becker2018} and GeV~\cite{Siyushev2017} centers. For SnV centers, spin initialization by optical pumping has been observed~\cite{Trusheim2020}. Another way of defining and driving a qubit could be to mix the (mostly) orthogonal spin components via an off-axis magnetic field, and use surface acoustic waves to drive phonon transitions between the orbital levels, with demonstrated Rabi frequencies $\sim$ 30 MHz~\cite{Maity2020}. Such an off-axis magnetic field, however, is at odds with achieving the high spin cyclicity needed for single shot spin state readout and high fidelity entanglement generation, necessitating a careful tradeoff~\cite{Nguyen2019detail}. For magnetic fields aligned with the color center's symmetry axis, readout fidelities above 99.99\% have been achieved~\cite{Sukachev2017,Nguyen2019overview}.

Tab.~\ref{tab:nvsnvproperties} also shows a comparison of the most important optical properties for both group-IV color centers and nitrogen-vacancy centers in diamond. To generate entanglement, only the fraction of coherent photons emitted in the zero-phonon-line (ZPL, for which no lattice vibrations assist the transition~\cite{Doherty2013}) can be used~\cite{Bernien2013}. Importantly, all quantified zero-phonon-line emission fractions for group-IV color centers are significantly larger than for the NV center, as the inversion symmetry of these emitters causes their excited states to have greater overlap with the optical ground state~\cite{Thiering2018}, resulting in a higher fraction of direct photon transitions~\cite{Franck1926,Condon1926}. While most recent quantum memory experiments with SiV centers have operated in dilution refrigerators (see below) in order to suppress thermal phonon population~\cite{Sukachev2017,Becker2018}, the SnV center with relatively high expected operating temperature requirements might be particularly promising for use as quantum network node. We note that the neutral charge state variants of group-IV color centers are also promising, as they potentially combine the favorable spin-photon interface of negative group-IV color centers with the excellent spin coherence times and easy qubit manipulation of negative nitrogen vacancy centers~\cite{Green2017,Rose2017,Thiering2018,Green2019,Zhang2020}. We do not discuss it extensively here due to its early stage of study.

\section{Enhancing the memory qubits} \label{Sec.:enhancing_memories}

We now discuss the state of the art and future research directions regarding the memory qubits of color centers in diamond. The most common isotope of carbon is $^{12}$C. However, the $^{13}$C isotope with natural abundance of about $1.1 \%$ carries a nuclear spin of 1/2. In the past decade, techniques have been developed (mostly on NV-based systems) to control these nuclear spins via the position-dependent hyperfine coupling. Universal nuclear spin control using electron decoupling sequences that are on resonance with a single $^{13}$C spin has been demonstrated ~\cite{Taminiau2014}. This has enabled the demonstration of coherence times above 10 seconds, and electron-nuclear gate fidelities $\sim 98 \%$ for individual $^{13}$C spins in NV-based systems~\cite{Kalb2017,Bradley2019}. Additionally, up to one memory qubit per communication node has been used in an NV-based quantum network setting to date~\cite{Kalb2017,Pompili2021}. For the case of SiV centers (and group-IV color centers in general), on the other hand, the spin-half nature of the system leads to a vanishing first-order sensitivity of decoupling sequences to individual $^{13}$C hyperfine parameters, thus requiring long decoupling times (and / or potentially off-axis magnetic fields that are at odds with a high spin cyclicity of optical transitions) to isolate out single nuclear spins. So far, electron-nuclear gate fidelities have been limited to $\sim 59 \%$ for SiV centers~\cite{Nguyen2019overview}; $^{13}$C nuclear spins close to group-IV color centers have not yet been used in quantum networks experiments.

\begin{figure}
\centering
\includegraphics[width=.4\textwidth]{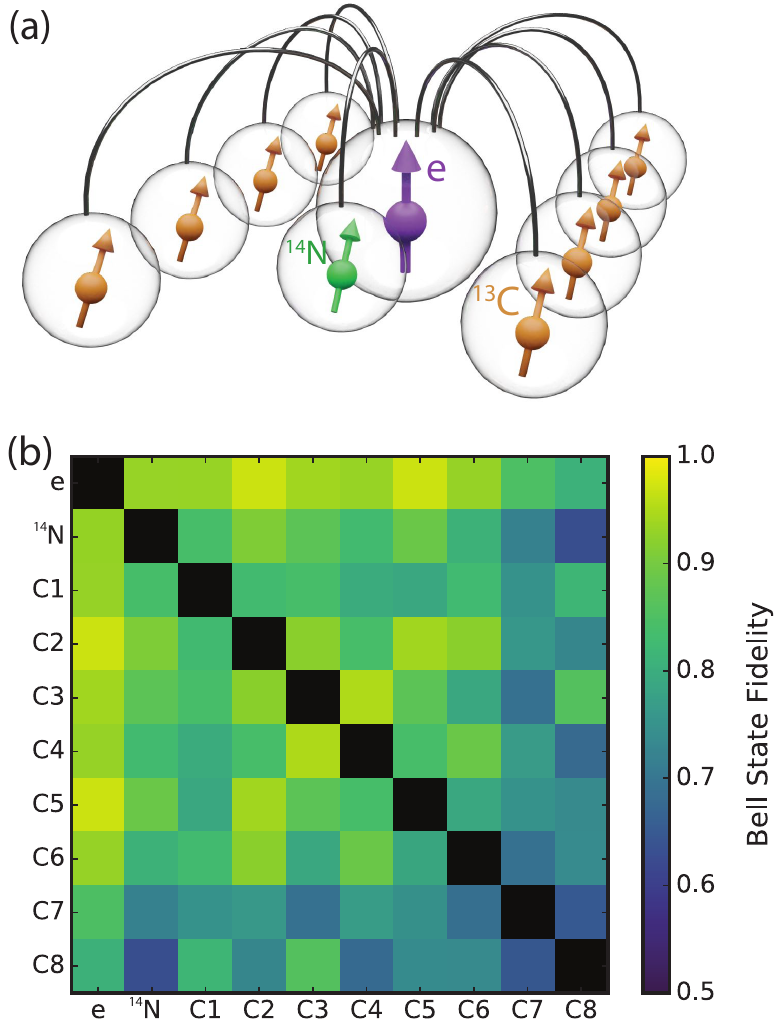}
\caption{State-of-the-art research for increasing memory qubit numbers and control for color centers in diamond. (a) Schematic overview of a local 10-qubit quantum register in diamond, formed from a nitrogen-vacancy center (purple), a native $^{14}$N nuclear spin, and 8 surrounding $^{13}$C nuclear spins. (b) Connection map of the spin register in (a), demonstrating entanglement generation between all pairs of qubits in the register. Figure adapted with permission from Ref.~\cite{Bradley2019}, in accordance with creativecommons.org/licenses/by/4.0/legalcode.} 
\label{fig:improving_communication_and_memory_qubits}
\end{figure}

A first area of research is thus to extend the number of available nuclear spin qubits per network node, as well as their control speeds and fidelities. Recently developed gate schemes based on interleaving radiofrequency (RF) driving of $^{13}$C nuclear spins (to control previously unaddressable qubits) with dynamical decoupling pulses (to decouple from the spin bath) have allowed the creation of a local quantum register of an NV communication qubit and up to 9 surrounding nuclear spin qubits [see \figref[top]{fig:improving_communication_and_memory_qubits}] that allow entangled state generation between all pairs of qubits [\figref[bottom]{fig:improving_communication_and_memory_qubits}], creation of a local 7-qubit GHZ state, as well as memory coherence times of single and two qubit states of over ten seconds~\cite{Bradley2019}. This result (achieved on a similar device as used for distant entanglement generation~\cite{Bernien2013,Kalb2017,Pompili2021}) shows that it is realistic to expect diamond-based quantum nodes containing many qubits in the near-term future. Such registers could enable scalable, modular quantum computation~\cite{Nickerson2014} and universal, fault-tolerant error correction~\cite{Taminiau2014,Waldherr2014,Cramer2016,Abobeih2021}. Additionally, we expect that RF driving of nuclear spins will overcome current limitations in manipulating nuclear spins for group-IV color centers, as the nuclear spin transition frequencies depend on the hyperfine parameters on first order in this case~\cite{Nguyen2019detail}. Overcoming issues related to sample heating and background amplifier noise could then also enable to address multiple nuclear spins simultaneously, thus reducing decoherence associated with long gate times~\cite{Bradley2019}. Additionally, cross-talk between nuclear spins, as well as unwanted coupling to other spins could be reduced when using the full information of the environment of a color center~\cite{Abobeih2019} (potentially acquired while involving automated techniques~\cite{Jung2021}) to simulate and tailor gate sequences for a specific spin environment, and computationally optimizing them for overall protocol fidelity. Another way of extending the number of controllable nuclear spin qubits per node is to also employ the nuclear spin of a nuclear spin containing color center atom isotope, e.g.~of $^{14}$N (nuclear spin of 1)~\cite{Pfaff2014} or $^{29}$Si (nuclear spin of 1/2)~\cite{Pingault2017}.

A second research direction is to increase the coherence time per available memory qubit under full network activity (in particular entanglement generation). The "always-on" nature of the hyperfine interaction of $^{13}$C nuclear spin memory qubits and the NV center has limited memory coherence in quantum networks to $\sim$ 500 entanglement generation attempts~\cite{Kalb2018,Pompili2021}. Uncontrolled electron dynamics, which result for instance from control infidelities and stochastic electron spin initialization during an entangling sequence, cause dephasing of the nuclear spin memory qubits~\cite{Kalb2018}. Techniques such as higher magnetic fields at the color center location can speed up gate times and shorten the entanglement generation element, thus reducing the time over which random phases can be picked up~\cite{Kalb2018,Pompili2021}. Additionally, decoupling pulses on the memory qubits that suppress quasi-static noise in the environment have shown initial promise to prolong nuclear memory qubit coherence under network activity for NV centers~\cite{Kalb2018}. Other promising routes to extend the nuclear spin memory coherence time involve reducing the color center's state-dependent coupling strength (the main dephasing channel), e.g.~by employing decoherence-protected subspaces (formed from two or more individual spins, or pairs of strongly-coupled spins that mostly cancel the state-dependent hyperfine interaction term)~\cite{Reiserer2016,Abobeih2018,Bartling2021}, and using isotopically purified samples for which weakly coupled $^{13}$C nuclear spin qubits can be controlled~\cite{Pfender2017,Pfender2019}). Other methods are to engineer systems of coupled defects (e.g.~involving a $^{13}$C nuclear spin qubit coupled to a P1 center in the vicinity of an NV center~\cite{Degen2020}, or to use the nitrogen nuclear spin of a second NV center (whose nitrogen nuclear spin is used as a memory) in proximity to the communicator NV center~\cite{Yamamoto2013,Degen_thesis_2021}.

\section{Enhancing the spin-photon interface} \label{Sec.:enhancing_optical_interface}

Optically-mediated entanglement generation with high rates requires quantum network nodes with efficient spin-photon interfaces. However, the relatively high refractive index of diamond ($\sim 2.4$) leads to significant total internal reflection at the diamond-air interface, which limits photon collection efficiencies. Entanglement rates are further reduced by the finite fraction of photons emitted into the zero-phonon line (ZPL) that can be used for entanglement generation (see Sec.~\ref{Sec.:Optial_and_spin_properties} above). Thus, techniques that can increase the fraction of collected photons at ZPL wavelengths, and / or increase the interaction of light with the emitter are required. Here, we discuss both strategies, as well as their suitability for enhancing the spin-photon interface of different diamond color centers, and the state of the art and future avenues for this research. We note that strong light-matter interactions also open the door to entangling schemes alternative from the photon-emission-based protocols discussed above~\cite{Cabrillo1999,Barrett2005}, e.g.~based on spin-dependent cavity reflections~\cite{Sorensen2003_reflection,Duan2004,Reiserer2015,Bhaskar2020,Chen2021}.

\subsection{Strategies for enhancing the spin-photon interface}

\begin{figure}
\includegraphics[width=.49\textwidth]{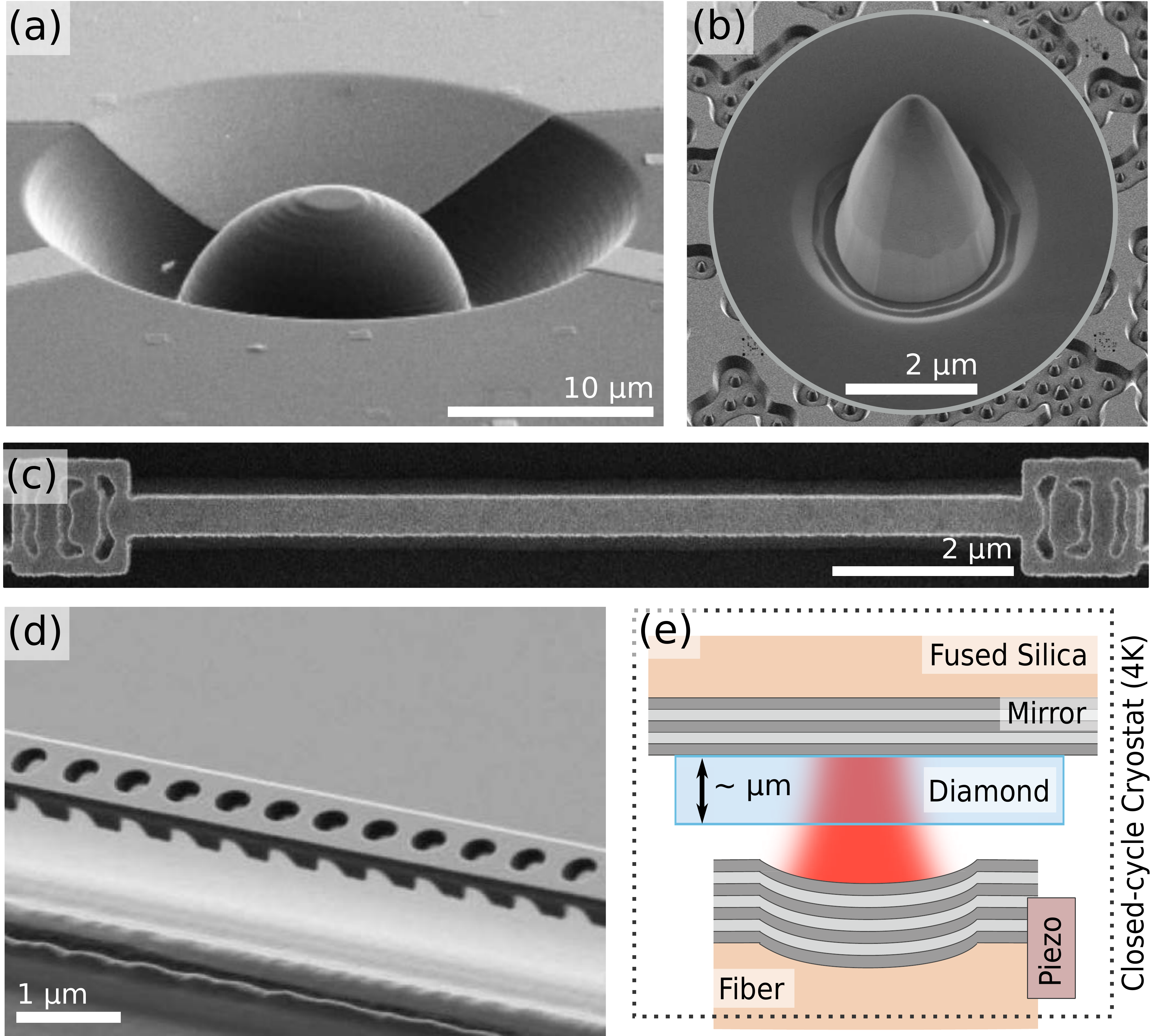}
\caption{Devices used to enhance the spin-photon interface of color centers in diamond. Solid immersion lenses~\cite{Robledo2011} (a), parabolic reflectors~\cite{Wan2017} (b, adapted with permission from Ref.~\cite{Wan2017}. Copyright (2017) American Chemical Society.), and waveguide (c, SEM image of device from the same chip used in Ref.~\cite{Rugar2020}) allow to enhance photon collection efficiencies from embedded color centers, compared to emission in a bulk host material. Photonic crystal~\cite{Burek2014} [d, Figure adapted with permission from Ref.~\cite{Burek2016b}, copyright by The Optical Society (OSA)] and open, tunable micro-cavities~\cite{Ruf2021} (e, Figure adapted with permission from Ref.~\cite{Ruf2021}, in accordance with creativecommons.org/licenses/by/4.0/legalcode) enhance collection efficiency, while modifying the nanophotonic environment at the same time. This allows to additionally enhance the number of photons emitted at a certain wavelength via the Purcell effect, compared to collection enhancement tools.}
\label{fig:collection_and_purcell_enhancement}
\end{figure}

The two most common strategies to enhance the spin-photon interface are by increasing the photon flux at the detector through improvements in photon collection, and through cavity or Purcell enhancement, which increases light-matter interaction and spontaneous emission rates (see Fig.~\ref{fig:collection_and_purcell_enhancement}). Here, we only give a brief overview of these methods, and refer to the many extensive diamond-color-center specific review articles that cover them in detail, see e.g. Refs.~\cite{Schroder2016,Johnson2017,Bradac2019,Janitz2020}.

\subsubsection{Collection efficiency enhancement}

A common method to enhance collection efficiency from color centers in diamond is to fabricate dome-shaped solid immersion lenses around the emitters [\figref[a]{fig:collection_and_purcell_enhancement}], which leads to a higher fraction of light being collected, as total internal reflection is avoided by the light striking the diamond surface at a perpendicular angle. These devices have been used to increase emission from NV~\cite{Hadden2010,Marseglia2011,Robledo2011,Jamali2014,Riedel2014}, SiV~\cite{Hepp2014,Rogers2014}, and GeV~\cite{Siyushev2017,DChen2019} centers; all diamond-based remote entanglement generation experiments reported to date have used these devices, with detection efficiencies of emitted photons of up to $\sim$ 10\%~\cite{Robledo2011}. 

A second class of collection efficiency enhancement methods is based on nanostructures that modify the far-field emission of a dipole emitter. These structures comprise different dimensions, designs, and supported wavelength ranges, and include parabolic reflectors~\cite{Wan2017,Hedrich2020} [\figref[b]{fig:collection_and_purcell_enhancement}], nanopillars~\cite{Maletinsky2012a}, nanowires~\cite{Hausmann2010,Babinec2010}, nanocones~\cite{Jeon2020}, gratings~\cite{Choy2013, Li2015circulargratings} and diamond waveguides~\cite{Burek2012,Momenzadeh2015, Shields2015,Mouradian2015,Patel2016,Sipahigil2016,Bhaskar2017,Rugar2020,Wan2020} [\figref[c]{fig:collection_and_purcell_enhancement}]. While they offer large-scale fabrication and higher collection efficiencies than solid immersion lenses (reported dipole-waveguide coupling efficiencies above 55\%~\cite{Hedrich2020,Wan2020}), the demands in nano-fabrication and color center placement are more stringent than for solid-immersion lenses, and emitters have to be brought in close proximity to nano-fabricated surfaces. This proximity to surfaces is believed to be the cause of optical instability for some diamond color centers, as discussed in detail below. Such instabilities make it challenging to produce indistinguishable photons from remote centers as required for entanglement generation ~\cite{Legero2003,Legero2006}.

\subsubsection{Purcell enhancement}

Another way of enhancing the spin-photon interface is to embed the dipole emitter inside an optical cavity, making use of the Purcell effect, described by a Purcell factor $F_P$. This factor scales with the ratio of the cavity quality factor, $Q$ (inversely proportional to the cavity's energy decay rate), to the cavity mode volume, $V$ (describing the cavity light field confinement)~\cite{Reiserer2015}. Briefly, a cavity enhances the density of states around its resonance frequency, such that the spontaneous emission of an embedded dipole emitter is increased when on resonance with the cavity following Fermi's golden rule. Additionally, the cavity funnels emitted light into a well-defined mode that can be readily matched e.g.~to that of optical fibers. A commonly used parameter to quantify the radiative emission enhancement by the cavity (defined as the ratio of the dipole-cavity emission, $\gamma_\textrm{cav}$, compared to the emission in a homogeneous medium, $\gamma_\textrm{hom}=\gamma_\textrm{rad}+\gamma_\textrm{nonrad}$) is the cooperativity, $C$, defined as 
\begin{equation}
   C =\frac{4g^2}{\kappa\gamma_\textrm{hom}} = \frac{\gamma_{cav}}{\gamma_{hom}} = F_P\beta_0\eta,
\end{equation}
where $g$ is the dipole-cavity coupling rate, $\kappa$ is the energy decay rate from the cavity, $\gamma_\textrm{rad}$ is the free-space radiative emitter decay rate, $\gamma_\textrm{nonrad}$ is the non-radiative emitter decay rate, $\beta_0$ is the fraction of photon emission from the emitter into the zero-phonon line (given by the Debye-Waller factor), and $\eta$ is the quantum efficiency (defined as the ratio of radiative to all decay). While this quantifies the efficiency of the spin-photon interface, it does not take into account the coherence of the emitter: in practice, solid state emitters often experience dephasing due to unwanted interactions with the environment of the solid state host, characterized by a pure dephasing rate $\gamma_\textrm{dep}$, which manifests in broadening of the optical transitions above the lifetime limited value~\cite{Robledo2010,Santori2010a,Riedel2017a,Jantzen2016,Nguyen2019overview,Rugar2020}. Therefore, it is useful to introduce the coherent cooperativity, $C_\textrm{coh}$, which quantifies the ratio of coherent decay into the cavity mode to undesired decay, and is given as~\cite{Borregaard2019,Janitz2020}
\begin{equation}
    C_\textrm{coh} = \frac{4g^2}{\kappa(\gamma_\textrm{hom}+\gamma_\textrm{dep})} = C\frac{\gamma_\textrm{hom}}{\gamma_\textrm{hom}+\gamma_\textrm{dep}}.
\end{equation}
$C_\textrm{coh}$ can be interpreted as the probability of coherent atom-photon interaction, and plays an important role in the fidelity and efficiency of many near-deterministic quantum protocols (which require $C_\textrm{coh} \gg 1$)~\cite{Borregaard2019}. 

To date, cavity enhancement of diamond color centers has been demonstrated for different implementations, including diamond-on-insulator whispering gallery mode resonators~\cite{Faraon2011,Hausmann2012} and nanodiamond plasmonic antennas~\cite{Bogdanov2020}. Another class of devices are based on the coupling of non-diamond resonators to color centers in nanodiamonds~\cite{Wolters2010,VanDerSar2011,Fehler2019,Fehler2020} or diamond films~\cite{Englund2010,Barclay2011,Gould2016}. Open and tunable micro-cavities geometries have also been demonstrated for embedded nanodiamonds~\cite{AlbrechtCoupling2013,Kaupp2013, Albrecht2014,Johnson2015,Kaupp2016,Benedikter2017} and thin diamond membranes~\cite{Riedel2017a,Jensen2019,Haussler2019,Salz2020,Ruf2021} [\figref[d]{fig:collection_and_purcell_enhancement}]. Steady progress in diamond nanofabrication has also led to all-diamond photonic crystal cavities~\cite{Faraon2012,Hausmann2013,Li2015,Riedrich-Moller2015,Sipahigil2016,Schroder2017,Evans2018,Jung2019a,Nguyen2019overview,Nguyen2019detail,Bhaskar2020}[\figref[e]{fig:collection_and_purcell_enhancement}], with demonstrated coherent cooperativities exceeding 100 for the case of the SiV center~\cite{Bhaskar2020}.

\vspace{0.8cm}

Quantum network applications also require the efficient coupling of light from these nanostructures to optical fibers. Different methods of coupling to fiber-based communication channels have been realized to date, including optimized free-space couplers with efficiencies up to $\sim 25\%$~\cite{Dory2019,Rugar2020}, notch couplers with efficiencies $\sim 1\%$~\cite{Sipahigil2016}, grating couplers with efficiencies $\sim 30\%$~\cite{Hausmann2012,Rath2013}, and double-~\cite{Khanaliloo2015a,Patel2016,Mitchell2019} and single-~\cite{Burek2017,Nguyen2019detail,Evans2018,Bhaskar2020} sided fiber-tapers with efficiencies of up to $\sim 90\%$.

\subsection{Looking forward: key challenges and potential solutions}

Having introduced different theoretical concepts to increase the spin-photon interface of color centers in diamond, we now discuss the experimental state of the art of such enhancement, highlighting key challenges and potential research directions.

\subsubsection{NV centers} 

While the nitrogen-vacancy (NV) center has been a workhorse of quantum network demonstrations with color centers in diamond, its low fraction of optical emission at the zero-phonon line (ZPL) frequency of 3\%, as well as limited collection efficiency from solid immersion lenses currently restrict local entanglement generation rates to <100 Hz~\cite{Humphreys2018,Pompili2021}. Due to their permanent electric dipole moment, NV centers experience large spectral shifts on short timescales caused by charge fluctuations in the environment when close to surfaces ($\sim$ microns)~\cite{Ruf2019}. As a result, despite reported Purcell factors of up to 70 in small mode-volume photonic crystal cavities~\cite{Faraon2012}, the regime of $C_\textrm{coh} \ge 1$ has remained out of reach. We note that the origins of the surface noise are still a subject of active research, and that recent experiments geared at understanding the effect and origins of surface noise on the NV center spin~\cite{Bluvstein2019}, and the local electrostatic environment of NV centers~\cite{Sangtawesin2019} show promise towards understanding the causes of surface charge noise. Combined with proposals for active optical driving to reduce spectral diffusion effects~\cite{Fotso2016}, this could revive the field of NV centers in nanophotonic structures in the future.

\begin{figure*}
    \centering
    \includegraphics[width=.65\textwidth]{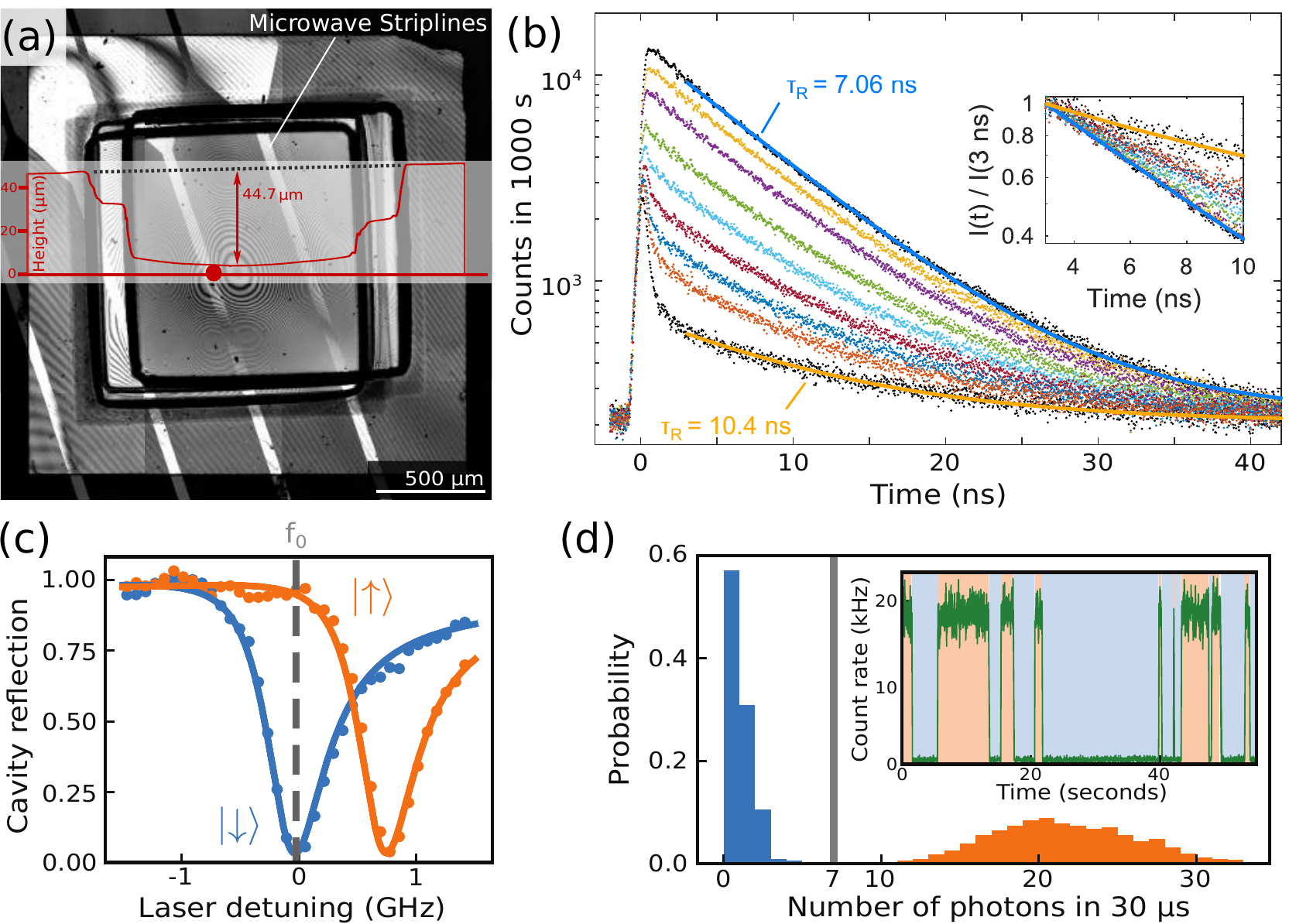}
    \caption{State-of-the-art research for enhancing the spin-photon interface of color centers in diamond for quantum networks applications. (a) Confocal microscope image of a typical \umdot-thin etched diamond sample containing NV centers with bulk-like optical properties, bonded via Van-der-Waals forces to a mirror containing microwave striplines. Location and direction of a sample height trace measurement is indicated with a red arrow. Such samples have recently enabled the first demonstration of resonant excitation and detection of coherent Purcell enhanced NV centers~\cite{Ruf2021}. Figure adapted with permission from Ref.~\cite{Ruf2019}, further permissions related to the material excerpted should be directed to the American Chemical Society. (b) NV center excited state decay in the ZPL for different emitter-cavity detunings for an NV center in an open micro-cavity under pulsed off-resonant excitation. Inset shows normalized decay curves. Figure adapted with permission from Ref.~\cite{Riedel2017a}, in accordance with creativecommons.org/licenses/by/4.0/legalcode. (c) Spin-state dependent reflection spectrum of a SiV center in a critically coupled photonic crystal cavity with $C_\textrm{coh}\sim100$. The spin-state dependent photon reflection with high collection efficiency enables spin-state readout with fidelity $>$ 99.9 \% in a single shot (d), and has recently enabled the first demonstration of memory-enhanced quantum communication~\cite{Bhaskar2020}. Figures (c) and (d) are adapted with permission from Ref.~\cite{Bhaskar2020}.}
    \label{fig:state_of_the_art_optical}
\end{figure*}

The challenge imposed by nearby surfaces can be circumvented by embedding microns-thin diamond membranes between two highly reflective mirrors~\cite{Janitz2015}. A recently reported fabrication technique has formed NV centers with bulk-like optical properties in \umdot-thin diamond membranes that can be embedded in such cavities via a combination of electron irradiation, high temperature annealing, and a tailored diamond etching sequence~\cite{Ruf2019} [\figref[a]{fig:state_of_the_art_optical}]. This has allowed for the first demonstration of resonant excitation and detection of optically coherent NV centers, Purcell enhanced by an optical cavity~\cite{Ruf2021}, as would be required for cavity-enhanced entanglement generation. However, the Purcell enhancement in this experiment was limited to $\sim3$, (mostly) due to cavity-length fluctuations $\sim$ 100 pm induced by a closed-cycle cryocooler. Previously, it has been shown (using off-resonant excitation) that Purcell enhancement factors of up to 30 (corresponding to 46\% of light emitted in the ZPL) can be achieved for such open micro-cavities in a liquid helium bath cryostat (featuring lower cavity length fluctuations)~\cite{Riedel2017a} [\figref[b]{fig:state_of_the_art_optical}]. Combined, these experiments demonstrate that there is a near-term path from the current $C_\textrm{coh}\sim 0.1$ towards $C_\textrm{coh} \sim 1$ and coherent photon collection efficiencies $\sim 10\%$. In addition, recent results suggest that it should be possible to reach the required pm-scale cavity length fluctuations even in closed-cycle cryostats~\cite{Merkel2020,Fontana2021}, which would remove the need for a liquid helium infrastructure at each quantum network node.

Another active area of research is to create shallow, stable NV centers~\cite{Lekavicius2019,Kasperczyk2020,Yurgens2021} at controlled locations; both the electron irradiation technique employed in Ref.~\cite{Ruf2021}, as well as recently introduced laser-writing techniques~\cite{Chen2019} can form optically coherent NV centers, but rely on native nitrogen in the sample to recombine with introduced lattice vacancies and thus miss precise control of the site of NV center formation. However, for maximal Purcell enhancement, NV centers should be positioned at an antinode of the cavity field. While such precision can be achieved using ion implantation, recent research has shown that NV centers created via this technique suffer from increased optical linewidths compared to NV centers formed during growth~\cite{VanDam2019}, even after extended high temperature treatments to restore the diamond lattice~\cite{Chu2014}. 

While it will remain challenging to achieve coherent cooperativites $\gg 1$ (and thereby enter the near-deterministic spin-photon interface regime~\cite{Borregaard2019}), the open cavity approach is projected to speed up current emission-based entanglement generation schemes by $\sim$ 2 orders of magnitude~\cite{Ruf2021}. This could allow for continuous deterministic entanglement generation (generating high fidelity entanglement faster than it is lost), and enable experiments such as the formation of a quantum repeater surpassing direct transmission~\cite{Rozpedek2019}, and device-independent quantum key distribution~\cite{Murta2019} using NV centers.

\subsubsection{Group-IV color centers}

The first-order electric field insensitivity of group-IV color centers in diamond~\cite{Sipahigil2014,Evans2016,Thiering2018,Muller2014,Iwasaki2015,Tchernij2017,Trusheim2020,Rugar2020,Trusheim2019} has enabled the demonstration of close-to-lifetime limited optical transitions even in heavily fabricated nanophotonic structures~\cite{Evans2016,Sipahigil2016,Bhaskar2017,Nguyen2019detail,Rugar2020, Wan2020}. To date, photonic crystal cavity coupling of group-IV color centers in diamond at low temperatures has been demonstrated for SiV~\cite{Sipahigil2016,Evans2018,Nguyen2019overview,Nguyen2019detail,Bhaskar2020} and SnV centers~\cite{Rugar2021}, with demonstrated $C_\textrm{coh} > 100$ for the case of the SiV center~\cite{Bhaskar2020}. This system allows for spin-state dependent photon reflection, enabling high fidelity single shot spin state readout with demonstrated fidelities $> 99.9 \%$ [see \figref[c,d]{fig:state_of_the_art_optical}]. This has recently enabled the first demonstration of memory-enhanced quantum communication~\cite{Bhaskar2020}.

A key requirement for many remote entanglement generation schemes is the ability to tune two group-IV color centers located on separate chips to a common resonance frequency. So far, tuning the emission frequency of group-IV emitters has been demonstrated using strain~\cite{Meesala2018,Sohn2018,Maity2018,Machielse2019,Wan2020}, electric fields~\cite{DeSantis2021,Aghaeimeibodi2021}, and Raman-type~\cite{Sipahigil2016,Sun2018} schemes (although the latter is only compatible with emission-based entanglement generation schemes), but only for single emitters, or several emitters in one structure. The first two tuning techniques deform the orbitals of the group-IV color centers, and thus the color's inversion symmetry is broken, leading to an observed increase in transition linewidths and spectral diffusion under applied external strain / electric field (potentially due to an increase in sensitivity to charge noise in the environment). We note that recent experiments indicate a larger tuning range of transitions for strain tuning (as compared to electric field tuning) for the same induced line broadening~\cite{Wan2020,DeSantis2021,Aghaeimeibodi2021}.  

While close-to-lifetime limited linewidths of group-IV color centers in nanophotonic structures have been observed, experiments still routinely show spectral shifts and charge instabilities, leading to broadening $\sim$ several lifetime-limited linewidths~\cite{Evans2018,Nguyen2019detail,Rugar2020, Wan2020}, as well as large local strain fields. While dynamic strain tuning can be used to suppress slow spectral diffusion (seconds timescale)~\cite{Machielse2019}, it is challenging to improve the homogeneous linewidth (sub-microsecond timescale) that enters the coherent cooperativity using this approach. Typically, emitters are created by high-energy implantation, followed by a high temperature annealing step to form group-IV vacancy centers and to reduce the effects of lattice damage from the implantation process~\cite{Chu2014}. However, there is evidence that even such high temperature treatments can not fully recover the original diamond lattice~\cite{VanDam2019}. A recently developed promising method to overcome this limitation employs low energy shallow ion implantation, combined with overgrowth of diamond material~\cite{Rugar2020overgrowth}. Another strategy could be to combine low-density ion implantation with electron irradiation, to reduce the amount of damage created in the lattice via the (heavy) ion implantation~\cite{McLellan2016}. Combined with controlled engineering of the diamond Fermi level~\cite{Collins2002,Rose2017,Murai2018}, these techniques could increase the quality and stability of group-IV color center optical transitions~\cite{Tchernij2017,Evans2018,Nguyen2019detail,Trusheim2019,Rugar2020, Wan2020}.

Another area of active development involves the design and fabrication of the nanophotonic structures. Despite recent progress~\cite{Schreck2017,Nelz2019}, growing high quality, thin film diamond on large scales is challenging, and (to the best of our knowledge) there is no wet processing technique that can etch bulk single-crystal diamond along its crystal planes, thus requiring either laborious thinning of diamond on a low index material, or sophisticated techniques to produce suspended, wavelength-thick diamond devices. While initial research focused on creating sub-\umdot-thin diamond films via selectively wet-etching a localized graphitized diamond layer~\cite{Parikh1992,Fairchild2008,Gaathon2013,Lee2014,Piracha2016,Piracha2016a} or etching down a super-polished thin diamond membrane~\cite{Faraon2012,Hausmann2013,Riedrich-Moller2014,Li2015,Cady2019}, these methods typically feature low device yield (as a consequence of challenges in device handling and initial material thickness variations). Thus, recent research has focused on fabricating structures directly in commercially available bulk diamond material via a combination of a hard mask and an angled diamond etch (leading to a triangular device cross-section)~\cite{Burek2012,Bayn2014,Burek2014,Sipahigil2016,Atikian2017,Zhang2018,Nguyen2019detail}, or a quasi-isotropic (dry) undercut etch, which selectively etches along certain diamond crystal planes (leading to rectangular device cross-section)~\cite{Khanaliloo2015,Khanaliloo2015b,Mouradian2017,Wan2018,Dory2019,Zheng2019,Wan2020}. Known limitations to these processes include mixing of TE- and TM-like modes that will ultimately limit device quality factors for triangular devices, as well as a relatively high bottom surface roughness for quasi-isotropic etched devices (although it is worthwhile to note that the fabrication parameter space is somewhat less explored for the latter technique). 

Currently, all listed fabrication methods routinely achieve photonic crystal cavity quality factors $\sim 10^4$, about two orders of magnitude lower than simulated values~\cite{Mouradian2017}. These deviations are caused by a combination of surface roughness~\cite{Nguyen2019detail}, non-uniform hole sizes~\cite{Rugar2021}, and deviations from the expected device cross-section~\cite{Wan2018}; we expect that an order-of-magnitude improvement in device quality factors is within reach upon further optimization. Furthermore, although photonic crystal nanocavities can be designed and fabricated to resonate at the ZPL frequency, they are especially sensitive to process variations, leading to a resonance wavelength spread of $\sim$ 5~nm across devices~\cite{Mouradian2017}. Nonetheless, this spread in frequency may be overcome by cavity tuning methods~\cite{Mosor2005,Faraon2012,Sipahigil2016}, which are in any case needed for the precise overlapping of the ZPL and cavity frequency. 

As nanofabrication methods are constantly refined, it is also likely that current methods of optimally aligning color centers within nanophotonic structures (to guarantee maximal overlap between the optical mode and the dipole emitter) can be further improved. Towards this end, various methods have already been demonstrated, including the targeted fabrication around pre-located centers~\cite{Wan2017}, or ion implantation into devices using either masks~\cite{Nguyen2019detail} or focused implantation~\cite{Sipahigil2016,Wan2020}. Compared to the SiV center for which $C_\textrm{coh}\sim 100$ has been demonstrated, GeV and SnV centers have intrinsically higher radiative efficiencies and thus potentially higher cooperativities, which may increase cavity-QED-based protocol fidelities (see discussion below).

We note that group-IV color centers in diamond are also suitable for integration into tuneable, open micro-cavities, as recently demonstrated for GeV~\cite{Jensen2019} and SiV~\cite{Haussler2019,Salz2020} centers at room temperature. It should be possible to achieve $C_\textrm{coh} \gg 1$ for these systems, potentially providing a viable alternative to the more fabrication-intensive nanostructures.

\section{Towards large-scale quantum networks} \label{Sec.:Large-scale integrated devices}

\begin{figure*}
\includegraphics[width=0.99\textwidth]{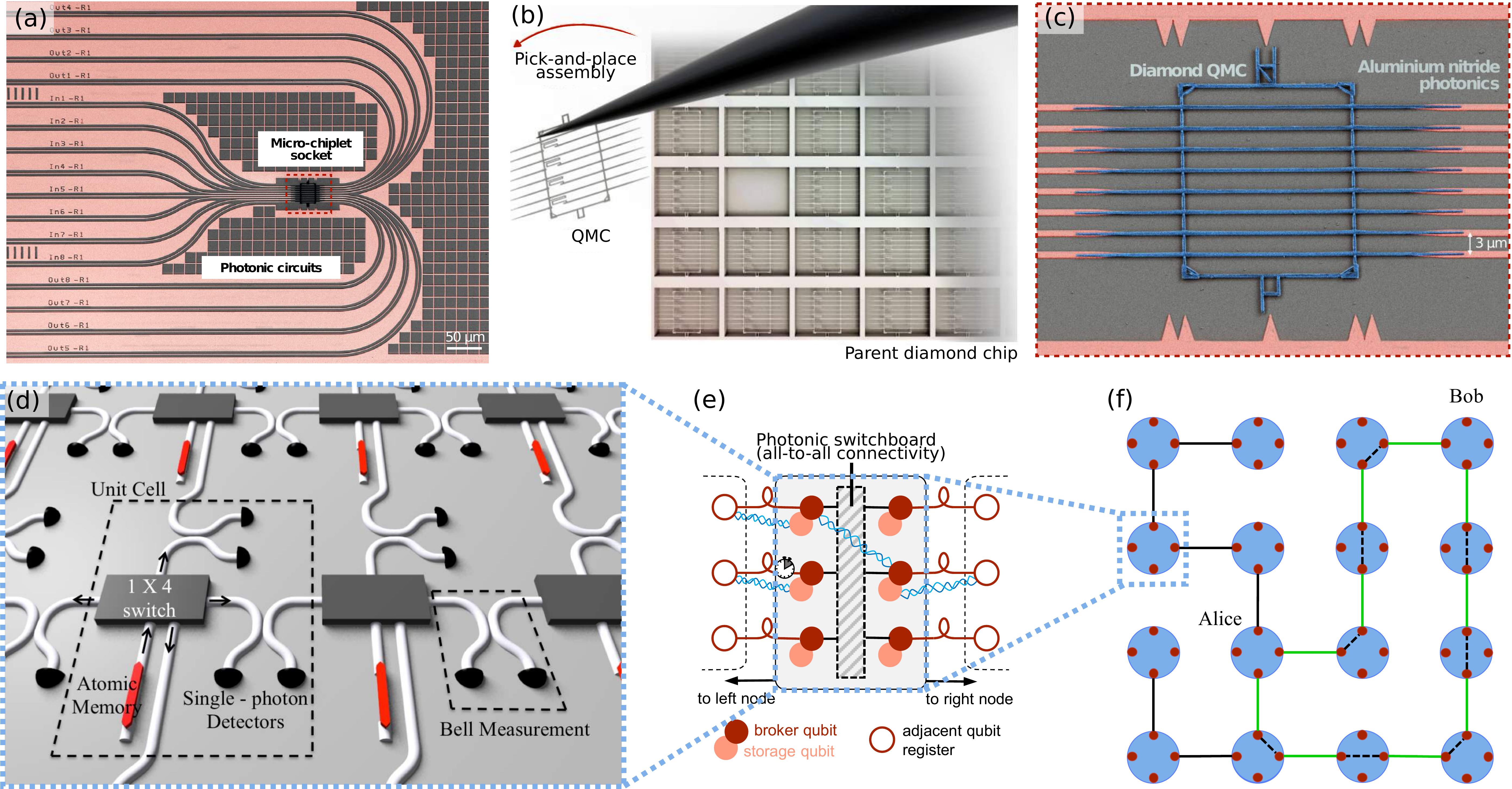}
\caption{Schematic of realized and envisioned architectures for large-scale hybrid integrated circuits enabling entanglement routing. (a) False-color SEM image of a AlN-on-sapphire photonic integrated circuit (PIC), containing a diamond quantum microchiplet [QMC, red dashed area, enlarged view in (c)]. These microchiplets with embedded color centers are pre-fabricated and -characterized in the diamond host material, and transferred onto the PIC using a pick-and-place technique (b), allowing to optimize PICs and microchiplets separately. This technique has been used to generate a defect-free array of 128 nanophotonic waveguides containing individually addressable quantum emitters, coupled to the PIC~\cite{Wan2020}. Images (a-c) adapted with permission from Ref.~\cite{Wan2020}. (d) Integrated multi-quantum memory node in a modular architecture~\cite{choi2019percolation}. Each unit cell (black dashed line) contains a diamond-based atomic memory (red), a fast, reconfigurable switching network allowing all-to-all connectivity (realized e.g.~via Mach-Zehnder interferometers), and photo-detectors. The atomic memory includes one communication and several memory qubits, with the communication qubit coupled to a diamond nanophotonic structure. Figure reprinted with permission from Ref.~\cite{choi2019percolation}, in accordance with creativecommons.org/licenses/by/4.0/legalcode.
(e) Entanglement routing in a quantum network. A multi-path routing algorithm (realized e.g.~via re-configurable PICs, dashed black box) can improve the entanglement generation rate (wiggled blue lines) between distant nodes (white circles), compared to a linear repeater topology~\cite{Lee2020}. This allows concurrent communication links with only local information used in each node (f, data and communication qubits depicted in light and dark red, respectively). Figure (e) adapted with permission from Ref.~\cite{Lee2020}. Figure (f) reprinted with permission from Ref.~\cite{Pant2019}, in accordance with creativecommons.org/licenses/by/4.0/legalcode.
}
\label{fig:scaling}
\end{figure*}

Having discussed the core quantum network components based on diamond color centers, we now turn to what is required to build these into a network capable of distributing entanglement over large distances at high rates.

First, a future quantum internet will likely make use of existing fiber infrastructure. This necessitates the matching of the photon wavelength from the network node to the telecommunication bands using quantum frequency conversion. Recently, it has been demonstrated that spin-photon entanglement can be preserved after frequency down-conversion of a $\sim 637$~nm photon entangled with the spin-state of the NV center to telecommunication wavelengths ($\sim 1588$ nm)~\cite{Tchebotareva2019}. Similar conversion techniques can be applied to other color centers and quantum systems~\cite{Krutyanskiy2019,Bock2018,Ikuta2018,Yu2020,DeGreve2012}. Future work will focus on increasing the system efficiency and potentially integrating on-chip conversion (see below).

Second, future quantum networks covering large distances will require many quantum repeater stations --– each with a large number of qubits --– for multiplexing, purification and error-correction. For example, 5-10 repeater stations with kilohertz entangled bit (ebit) rates need a total of $O(10^8)$ data qubits to reach mega-ebits per second communication \cite{muralidharan2016}. Although color center devices are already produced using standard nanofabrication techniques, the needed scale for high-rate, high-fidelity networks would require large-scale manufacturing of quantum nodes. This task not only entails qubit and device production but also the packaging of efficient optical and microwave signals to and from many independent color centers at once.

Optical technologies such as photonic integrated circuits (PICs) may play an important role in addressing these challenges. Their programmability~\cite{Bogaerts2020} and access to a large number of spatial modes~\cite{Harris2018} are especially of interest to quantum network applications. Similar to their bulk optics counterparts, PICs comprise of low-loss on-chip components, such as waveguides, filters and switches. References~\cite{Bogaerts2020,Blumenthal2018,Harris2018,Zhu2021} review the device concepts and state of the art of PIC technologies that may be relevant to quantum network applications. Driven by foundry adoption, as well as new frontiers in data communication~\cite{Sun2015,Atabaki2018}, photonic processors~\cite{Wetzstein2020,Shastri2021} and optical quantum computing~\cite{Wang2020}, integrated photonic circuits have advanced dramatically in manufacturability and complexity over the last decade. The first notable feature of PICs for quantum technologies is their compact footprint, which not only promotes dense integration, but also reduces phase errors in quantum interference of photons~\cite{Bogaerts2020}. Next, phase modulators in PICs can implement on-chip switches for routing photons within an optical network~\cite{Bogaerts2020}, and material nonlinearities can be used for efficient frequency conversion between visible and telecommunication photons~\cite{Wang2018}. Finally, multi-channel optical access can be accomplished using standard fiber arrays, and electrical packaging for large-scale control of color centers can be achieved using potential metal layers in a PIC stack. 

Photonic circuits in diamond have been previously demonstrated for opto-mechanics~\cite{Rath2013} and nonlinear optics~\cite{Hausmann2014}. Gallium phosphide-on-diamond photonics have also been used to route the emission from NV centers~\cite{Gould2016}. However, the non-deterministic creation and integration of color centers in devices, as well as the absence of single-crystal diamond wafers have limited the scale of diamond integrated photonics (see Sec.~\ref{Sec.:enhancing_optical_interface}). One way to combine the functionalities and performance of industry-leading photonic circuits with diamond is through heterogeneous integration of diamond with other material systems. Also known as "hybrid photonics"~\cite{Kim2020,Elshaari2020}, this approach is akin to modern integrated circuits in that discrete chips are separately optimized and fabricated and then populated into a larger circuit board. Examples of heterogeneous integration include modulators and lasers in silicon photonics~\cite{Komljenovic2018}. In the context of quantum photonics, recent successes include quantum emitters~\cite{Kim2020,Elshaari2020} and single-photon detectors~\cite{Najafi2015}, which are otherwise difficult to achieve in a single material platform with high performance.

Heterogeneous integration of diamond color centers with PICs circumvents the yield issues associated with all-diamond architectures, thereby allowing potentially many addressable qubits within a single chip. A recent result showing the large-scale integration of color centers in diamond with hybrid photonic integrated circuits~\cite{Wan2020} is shown in \figref[a-c]{fig:scaling}. Diamond quantum micro-chiplets, each consisting of 8 diamond nanophotonic waveguides with at least one addressable group-IV color center, are integrated with PICs based on aluminum nitride [\figref[c]{fig:scaling}]. After fabrication using the quasi-isotropic undercut technique (see Sec.~\ref{Sec.:enhancing_optical_interface}), a total of 16 chiplets numbering to 128 waveguides were positioned with sub-microns accuracy on the PIC [\figref[a]{fig:scaling}] using a pick-and-place technique [\figref[b]{fig:scaling}]. The coupling of the color center to the waveguide can be as high as 55 \%, and the diamond-PIC coupling and PIC-to-optical fiber coupling are reported to be 34 \% and 11 \%, respectively. In addition, electrodes in this hybrid PIC enabled the in-situ tuning of optical transition frequencies within this integrated device architecture~\cite{Wan2020}. The availability of multiple color centers per waveguide potentially allows for spectral multiplexing~\cite{Bersin2019, Chen2020}, which is a hardware-efficient path to multiplying the total number of qubits to $N_s \times N_f$, where $N_s$ and $N_f$ are the number of spatial and frequency channels, respectively.

With further improvements in waveguide-emitter coupling as well as diamond-nanocavity integration with PICs, such a hybrid architecture could become an important building block of future quantum network nodes. Looking forward, we expect future developments to also focus on integrating chip components for color center technologies, such as CMOS-integrated microwave electronics for spin control~\cite{Kim2019}, on-chip beamsplitters for photon quantum interference~\cite{Wang2020}, optical switches for channel connectivity~\cite{Bogaerts2020}, single-photon detectors~\cite{Reithmaier2015,Najafi2015,Schwartz2018,Gyger2021} for heralded entanglement, and quantum frequency conversion~\cite{Guo2016,Jankowski2020} or frequency tuning~\cite{Hu2020} on the same microphotonic platform. While these functionalities have been realized in other PIC platforms, a key challenge is to bring together various technologies and materials for implementing a chip-based node. Here, the advances in heterogeneous integration will be critical for controlling and deploying a large number of such chips. 

As an example of such a chip, \figref[d]{fig:scaling} shows a modular architecture with many unit cells per network node~\cite{choi2019percolation}. Each unit cell consists of atomic memories, photonic switches, and single-photon detectors. The atomic memories include communication qubits for inter-cell entanglement and memory qubits for long-term storage and intracell information processing. The switching network selects one of the adjacent cells, and the photo-detectors herald successful entanglement events between cells. Such a photonic architecture could allow for entanglement routing within a chip-based node, potentially boosting the communication rate~\cite{Duan2004,Lee2020}. By establishing connectivity between unit cells using optical switches, the resulting network is scalable because adding nodes does not require any modification in the existing network.

Once multiple long-lived memory qubits that can be controlled with high fidelity are available per quantum network node, the entanglement fidelity can be greatly improved through entanglement distillation (purification), where low fidelity entangled pairs are employed as a resource and transformed to a smaller number of high fidelity pairs after local operations and classical communication~\cite{Bennett1996,Deutsch1996}. Such a scheme has recently been demonstrated in a proof-of-principle experiment between two distant NV centers that each have access to an additional $^{13}$C memory~\cite{Kalb2017}. While such distillation requires additional time resources as a result from two-way-classical-communication, quantum error correction~\cite{Jiang2009} can achieve a more favorable key rate per qubit scaling\cite{muralidharan2016}. Such error correction can be performed if the operational error is smaller than the threshold of a given code (e.g.~$\approx 0.189$ for the Calderbank-Shor-Stean (CSS) code~\cite{Poulin2006}). If the photon-loss of a link is sufficiently small, heralded entanglement generation can also be replaced with error correction, allowing to move beyond two-way signalling associated with the heralding~\cite{Borregaard2020}. In addition, multi-qubit network nodes can be operated in a way that outperforms more simple linear repeater schemes. For example, \figref[f]{fig:scaling} shows a 2D network with nearest-neighbor connectivity that allows multi-path entanglement routing~\cite{Pant2019}. Crucially, the algorithm only requires local knowledge about which of the entanglement generation attempts with its nearest neighbor succeeded while still achieving faster key distribution than a linear repeater chain~\cite{Pant2019}.

\section{Conclusions} 
\label{Sec.:Conclusions}

We have provided a perspective on the emerging field of quantum networks based on diamond color centers. Diamond color centers already define the state of the art in multi-node entanglement-based networks~\cite{Pompili2021} and in memory-enhanced quantum key distribution~\cite{Bhaskar2020}. We expect the next years to see rapid progress on photonic interfaces and integration of color centers, paving the way for first experiments on long-distance quantum links. From the basic building blocks, larger-scale devices will be designed and constructed. Control layers of higher abstraction -- akin to the current Internet -- are currently being developed~\cite{Dahlberg2019,Pirker2019}.

A future functional quantum network will support many interesting applications, such as distributed quantum computing~\cite{Cuomo2020}, accessing a quantum server in the cloud with full privacy~\cite{Broadbent2009} and stabilizing quantum clocks~\cite{Hodges2013,Komar2014}. Color centers in diamond may play an essential role in these networks, as the ``satisfactory repeater'', or perhaps the ``transistor of the quantum age.''

\section*{Acknowledgements} 

We thank Ben Dixon, Mihir Bhaskar, Eric Bersin, Johannes Borregaard, Tim Taminiau, Matteo Pompili, and Matteo Pasini for feedback on the manuscript, Hans Beukers for the calculations of operation temperatures of color centers, and Matteo Pompili, Mihir Bhaskar, Christian Nguyen, Daniel Riedel, Alison Rugar, Shahriar Aghaeimeibodi, Conor Bradley, Mihir Pant, Yuan Lee, Michael Burek and Cleaven Chia for providing originals of figures. D.E.~acknowledges support from Brookhaven National Laboratory, which is supported by the U.S.~Department of Energy, Office of Basic Energy Sciences, under Contract No.~DE-SC0012704. N.W.~acknowledges support from the MITRE Quantum Moonshot initiative. H.C.~acknowledges the Claude E.~Shannon Fellowship and Samsung Scholarship. R.H.~and M.R.~acknowledge financial support from the EU Flagship on Quantum Technologies through the project Quantum Internet Alliance, from the Netherlands Organisation for Scientific Research (NWO) through a VICI grant and the Zwaartekracht program Quantum Software Consortium, and the European Research Council (ERC) through a Consolidator Grant.

\bibliography{jap}

\end{document}